\documentclass{aa}  
\usepackage{graphicx}
\usepackage{txfonts}
\usepackage[figuresright]{rotating}
\def\GS{\object{GS~1826$-$238~}}
\def\SAX{{\em BeppoSAX}~}
\def\Chandra{{\em Chandra}~}
\def\XTE{{\em RXTE}~}
\def\INTEGRAL{{\em INTEGRAL}~}
\def\XSPEC{{\sc Xspec}~}
\def\ergcms{{\rm erg~cm$^{-2}$s$^{-1}$}}
\def\CompTB{{\sc comptb}~}
\def\CompTT{{\sc comptt}~}
\begin{document}
   \title{Wide band observations of the X-ray burster \GS}

   \subtitle{}

   \author{M. Cocchi\inst{1}\thanks{e-mail: Massimo.Cocchi@iasf-roma.inaf.it}
          \and
           R. Farinelli\inst{2}
          \and
           A. Paizis\inst{3}
          \and
           L. Titarchuk\inst{2,4,5,6}
          }

   \offprints{M. Cocchi}

   \institute{INAF, Istituto di Astrofisica Spaziale e Fisica Cosmica sez. di Roma,
              via Fosso del Cavaliere, 100, 00133 Roma (Italy)
              \and
              Dipartimento di Fisica, Universit\'a di Ferrara (Italy)
              \and
              INAF, Istituto di Astrofisica Spaziale e Fisica Cosmica, sez. di Milano (Italy)
              \and
              Center for Earth Observing and Space Research, George Mason University, Fairfax, VA (USA)
              \and
              High Energy Space Environment Branch, US Naval Reserch Laboratory, Washington, DC (USA)
              \and
              Gravitational Astrophysics Laboratory, NASA Goddard Space Flight Center, Greenbelt, MD (USA)
             }

  \date{Received ?; accepted ?}


  \abstract
   {
    \GS is a well-studied X-ray bursting neutron star in a low mass binary system.
    Thermal Comptonisation by a hot electron cloud ($kT_{e} \sim 20$ keV) is a widely accepted mechanism accounting for its high energy emission, while the nature of most of its soft X-ray output is not completely understood. A further low energy component is typically needed to model the observed spectra: pure blackbody and Comptonisation-modified blackbody radiation by a lower temperature (a few keV) electron plasma were suggested to explain the low energy data.
   }
   {
    In order to better characterise the nature of the low energy emission and
    the bolometric output of the source, the steady emission of \GS is studied
    by means of sensitive, broad band (X to soft Gamma-rays) measurements obtained by the \INTEGRAL observatory.  
   }
   {
   In this data analysis, the newly developed, up-to-date Comptonisation model
   \CompTB is applied for the first time to study effectively the low-hard state variability of a low-luminosity neutron star in a low-mass X-ray binary system.
Three joint ISGRI/JEM-X data sets (two from observations performed in 2003 and one from 2006) were analysed.
   }
   {
We confirm that the 3--200 keV emission of \GS is characterised by Comptonisation of soft seed photons by a hot electron plasma. 
A single spectral component is sufficient to model the observed spectra. At lower energies, no direct blackbody emission is observed and there is no need to postulate a low temperature Compton region.
Compared to the 2003 measurements, the plasma temperature decreased from $\sim 20$ to $\sim 14$ keV in 2006, together with the seed photons temperature.  
The source intensity was also found to be $\sim 30\%$ lower in 2006, whilst the average recurrence frequency of the X-ray bursts significantly increased. 
Possible explanations for this apparent deviation from the typical limit-cycle behaviour of this burster are discussed.
   }


   \maketitle
%

\section{Introduction}

The persistent X-ray source \GS was discovered during a manoeuvre of the {\em
Ginga} satellite in 1988 (\cite{Makino}). 
Early band-limited observations of this object were performed by {\em TTM} (\cite{Zand92}), and {\em ROSAT} (\cite{Barret95}) at low energies and by {\em CGRO-OSSE} (\cite{Strickman}), in the hard X-rays.
An optical counterpart in a low inclination system and with a likely orbital period of 2.1 h was reported (\cite{Homer}).
The detection of type I X--ray bursts (\cite{Ubertini97}) by \SAX ultimately
established that the source is a neutron
star (NS) in a low mass binary system (LMXB). X--ray bursts were found to
recur every few hours in a quasi-periodical fashion (\cite{Ubertini99}). What
makes this burster unique among its class is the remarkable stability of the
recurrence over a time-scale of several years (\cite{Cocchi}). Bursting behaviour
is clearly related to the persistent emission (\cite{Cornelisse}), the event
frequency increasing with the measured X-ray flux. {\em RXTE} data permitted more accurate studies of the burst recurrence properties (see e.g. \cite{Galloway}), and a possible 611 Hz oscillation in the burst decay was also reported (\cite{Thompson}).
The lack of evidence for photospheric radius expansion during the bursts
  (\cite{Zand}) and general considerations on the optical counterpart (LMXB,
  short orbital period) limit the value of the distance to the source in the
  $\sim 4-9$ kpc range (\cite{Barret}).

The history of sensitive, wide band measurements of the steady emission of \GS
includes results from \SAX (\cite{Zand, DelSordo}), {\em RXTE} (\cite
{Barret}), {\em RXTE} plus {\em Chandra} (\cite{Thompson}), and {\em RXTE}
plus {\em XMM} (\cite{Thompson2}). Since the \SAX era, the source is always
detected in a hard (low) spectral state, its high energy emission extending
well in excess of 100 keV. A simple, empirical, exponentially-cutoff power law
model with $E_{c} \sim$ 50 keV yields acceptable fits to the high energy data
(e.g. \cite{DelSordo}) but whenever more physical models are considered, the
spectra are generally well described by a Comptonisation scenario, where soft
seed photons are up-scattered to high energies by a hot electron plasma
($kT_{e} \sim 20$ keV). Along with Comptonisation, either a low energy
blackbody (BB) spectral component (\cite{Zand}) or a low-temperature Compton
component (\cite{Thompson}) are included to fit the data.  The detection of a
$\sim 6$ keV iron line was reported by \cite{Barret}, who also included reflection to fit their data.
Evidence for transient soft blackbody emission, possibly related to irregularities in the time behaviour of the X-ray bursts, has recently reported by Thompson et al. (2008). 

We present wide band spectral analysis of \GS performed by \INTEGRAL data. In the next section, observation and data analysis details will be given. In Sect. 3 we present and discuss our  results along with our best fit models, while the implications of these results and their relation to the current knowledge of this peculiar X-ray source are debated in Sect. 4.

%
\begin{table}
\begin{minipage}[t]{\columnwidth}
\caption{Observation log.}
\label{table:1}
\centering
\renewcommand{\footnoterule}{}  
\begin{tabular}{l c c c}        
\hline\hline                 
\INTEGRAL revolution    & 061--064 & 119--122 & 495 \\  
\hline
obs. year  & 2003 & 2003 & 2006 \\
obs. day   & Apr 15--23 & Oct 06--15 & Nov 2 \\
JEM-X live time (ks)      & 21.9    & 32.8     &  49.9 \\
ISGRI live time (ks)      & 134.4   & 206.5    &  50.3 \\
\# of bursts (JEM-X)      &  3      &   5      &   4   \\
average burst delay (h)   &  3.79   &  3.77    &  3.40 \\
\\
ASM intensity (mCrab)~\footnote{average during the \INTEGRAL observations,
  2--10 keV.}  & $37\pm 7$ & $40 \pm 8$ & $25\pm 4$ \\
\hline\hline 
\end{tabular}
\end{minipage}
\end{table}

\section{Observations and data analysis}

The public database\footnote{http://isdc.unige.ch/index.cgi?Data+info} of \INTEGRAL (\cite{Integral}) includes a big set of useful pointings of GS~1826$-$238, thanks to the observing policy of the telescope which includes periodic scans of the Galactic plane and the Galactic bulge region (\GS is only $\sim 10$ degrees from SgrA*).

Detailed analysis of the whole \INTEGRAL \GS public database, focused on the source long term monitoring, is to be presented elsewhere (Cocchi et al., in prep.).
The present work, focused on wide band spectroscopy of the persistent emission, relies on data obtained by the imaging detectors JEM-X (\cite{JemX}) and IBIS/ISGRI (\cite{Isgri}, hereafter ISGRI), which operate in the 2--30 keV and 17--300 keV passbands, respectively. 

Being the source variable, though always observed in a low--hard spectral
state, homogeneous data sets are needed: this means that one has to choose
simultaneuos JEM-X+ISGRI pointings of sufficient length to get good statistics and as close in time as 
possible.
Due to the different field of view of the two detectors and the typical
dithering observing pattern of {\em INTEGRAL}, JEM-X live time is considerably less than the ISGRI one. 

Taking into account the above limitations, we selected three data sets, the
first one belonging to the satellite 
revolutions 061--064 (visibility window of Spring 2003), the second one to the
revolutions 119--122 (Autumn 2003), and the third one to revolution 495 (Autumn 2006). 
Together, the three sets account for $104.6$ ks and $391.2$ ks live time (JEM-X and ISGRI, respectively).
Analysis of {\em RXTE}/ASM monitoring data of \GS obtained in the same epochs
of our \INTEGRAL observations indicate that the 2--10 keV source intensity significantly varied from 2003 to 2006, being $\sim 40$ mCrab in 2003 and only $\sim 25$ mCrab in Autumn 2006 (see Table 1). This is confirmed with higher accuracy and on a wider energy band by the present measurements (see Sect. 3).

In order to avoid possible low-energy spectral contamination from the X-ray burst emission,
the time intervals of the observed bursts, each event typically lasting $\sim
100$ s in JEM-X data, were removed from the good time of the data sets.
Details about the analysed data sets such as live time, number of bursts
detected and inferred average burst periodicities, are summarised in Table 1.
 
JEM-X data were reduced and extracted by the standard \INTEGRAL off-line analysis (OSA) pipeline\footnote{http://isdc.unige.ch/index.cgi?Soft+info} release 7.0.
Presence of systematic effects at the lowest energies suggested a discarding
of the spectral bins in the 2--3 keV range. 
ISGRI data were reduced by the standard OSA 6.0 pipeline, while spectral extraction was performed by a dedicated 
software  based on cross correlation, which is freely available to the community\footnote{http://www.ifc.inaf.it/$\sim$ferrigno/INTEGRALsoftware.html}.
The effectiveness and the limitations of the software in terms of calibration
and systematics have already been discussed elsewhere (\cite{Segreto};~\cite{Mineo}).  
Finally, the joint JEM-X/ISGRI spectra were analysed and best-fitted by the standard software tools of the \XSPEC package\footnote{http://heasarc.nasa.gov/docs/xanadu/xspec/index.html}, version 12.

Due to cross calibration uncertainties between the two instruments,  
whose effects are expected to vary with the observation epoch$^{2}$,  
a constant normalisation factor was included in each fit.
We kept ISGRI as a reference, allowing the JEM-X constant to vary. 
No systematics, accounting for possible instrumental miscalibration, had to be included for both JEM-X and ISGRI.

Our data cannot constrain the low energy absorption along the line of sight, so the $N_{\rm H}$ value, modeled by Wisconsin absorber in \XSPEC, was kept fixed to $3\times 10^{21} \rm H~cm^{-2}$, which is the best determined value from recent joint XMM-RXTE measurements (\cite{Thompson2}). 
Slightly different values, $5\times 10^{21} \rm H~cm^{-2}$ and $1.5\times
  10^{21} \rm H~cm^{-2}$ were determined respectively by previous {\em ROSAT} and \SAX measurements (\cite{Barret95}; \cite{Zand}). All the values are of the same order of magnitude of the interpolated one (\cite{Dickey}) that can be derived by {\sc nh}, a widely used ftool\footnote{http://heasarc.nasa.gov/docs/software/ftools/ftools\_menu.html}.
We already verified that these slight differences in the $N_{\rm H}$ value do not affect the parameters of our best fits and the possible undetection of further soft components (see Sect. 3).

%
\begin{table}
\begin{minipage}[t]{\columnwidth}
\caption{Spectroscopic results.}
\label{table:2}
\centering
\renewcommand{\footnoterule}{}  
\begin{tabular}{l c c c}        
\hline\hline                 
data set (revolution)  & 061--064 & 119--122 & 495 \\  
\hline
\multicolumn{4}{c}{\sc wabs*CompTB~\footnote{name of the Xspec model. n$_{H}$ fixed to $3.0\times 10^{21}\rm{H~cm}^{-2}$ in all fits. All errors are 90\% probability.}}     \\ 
$k\rm{T}_{s}$~(keV)               & $1.06~^{+0.07}_{-0.06}$ & $0.95 \pm 0.05$         & $0.45~^{+0.09}_{-0.14}$ \\
$\alpha$ index                       & $1.05~^{+0.04}_{-0.03}$ & $1.03 \pm 0.03$         & $0.85~^{+0.02}_{-0.01}$ \\
$k\rm{T}_{e}$~(keV)                  & $20.1~^{+1.7}_{-1.2}$   & $19.4~^{+1.0}_{-0.9}$   & $13.5 \pm 0.8$          \\
const.~\footnote{JEM-X/ISGRI normalisation ratio (see Sect. 2).}  & $0.671~^{+0.028}_{-0.032}$ & $0.732~^{+0.027}_{-0.025}$ & $0.932~^{+0.029}_{-0.037}$ \\
$\rm{red.} ~\chi^{2}$ (d.o.f.)       &  1.076 (360)            &  1.092 (361)            &  1.094 (344)            \\ 
flux 2--20 keV~\footnote{all fluxes in \ergcms.}  & $1.11\times 10^{-9}$ & $1.34\times 10^{-9}$ & $1.02\times 10^{-9}$ \\
flux 20--200 keV                     & $1.23\times 10^{-9}$    & $1.34\times 10^{-9}$    & $0.80\times 10^{-9}$    \\
flux 2--200 keV                      & $2.34\times 10^{-9}$    & $2.68\times
10^{-9}$    & $1.82\times 10^{-9}$    \\
luminosity~\footnote{unabsorbed, 0.1-200 keV, assumed distance of 7 kpc}
  ({\rm erg~s$^{-1}$})  & $1.76\times 10^{37}$    & $1.92\times10^{37}$    &  $1.42\times 10^{37}$   \\  
$\tau$~\footnote{derived parameter (see Sect. 3) }     & $3.4$    & $3.5$                   & $5.3$ \\
$CAF^{~d}$                                           & $3.8$    & $3.9$                   &$11.5$ \\
\hline
\multicolumn{4}{c}{\sc wabs*CompTT} \\
$k\rm{T}_{0}$~(keV)               & $1.02 \pm 0.06$         & $0.91 \pm 0.05$         & $0.33 \pm 0.04$  \\
$k\rm{T}_{e}$~(keV)                  & $21.5~^{+2.2}_{-1.7}$   & $20.0~^{+1.2}_{-1.0}$   & $14.6~^{+1.3}_{-1.1}$   \\
$\tau$~\footnote{assumed spherical geometry}     & $3.81~^{+0.33}_{-0.36}$ & $4.12~^{+0.23}_{-0.24}$ & $5.55~^{+0.33}_{-0.35}$ \\
const. 			             & $0.680~^{+0.031}_{-0.029}$ & $0.756~^{+0.026}_{-0.025}$ & $0.900~^{+0.043}_{-0.041}$ \\
$\rm{red.} ~\chi^{2}$ (d.o.f.)       &  1.064 (360)            &  1.081 (361)            &  1.070 (344)            \\
flux 2--200 keV                      & $2.34\times 10^{-9}$    & $2.68\times 10^{-9}$    & $1.85\times 10^{-9}$    \\
\hline
\multicolumn{4}{c}{\sc wabs*cutoffpl} \\
E$_{\rm cut}$~(keV)            & $51.9~^{+4.0}_{-3.5}$      & $55.9~^{+3.2}_{-2.9}$ & $49.0~^{+6.1}_{-5.1}$ \\
$\Gamma$                       & $1.57 \pm 0.05$            & $1.62 \pm 0.03$       & $1.62 \pm 0.04$       \\
const. 				 & $0.784~^{+0.035}_{-0.034}$ & $0.829~^{+0.028}_{-0.027}$ & $0.825~^{+0.038}_{-0.037}$ \\
$\rm{red.} ~\chi^{2}$ (d.o.f.) &  1.022 (361)               &  1.065 (362)          &  1.001 (345)          \\
flux 2--200 keV                & $2.46\times 10^{-9}$       & $2.83\times 10^{-9}$  & $1.91\times 10^{-9}$  \\ 
\hline\hline 
\end{tabular}
\end{minipage}
\end{table}

   \begin{figure*}
   \centering
    \includegraphics[angle=90,width=16cm]{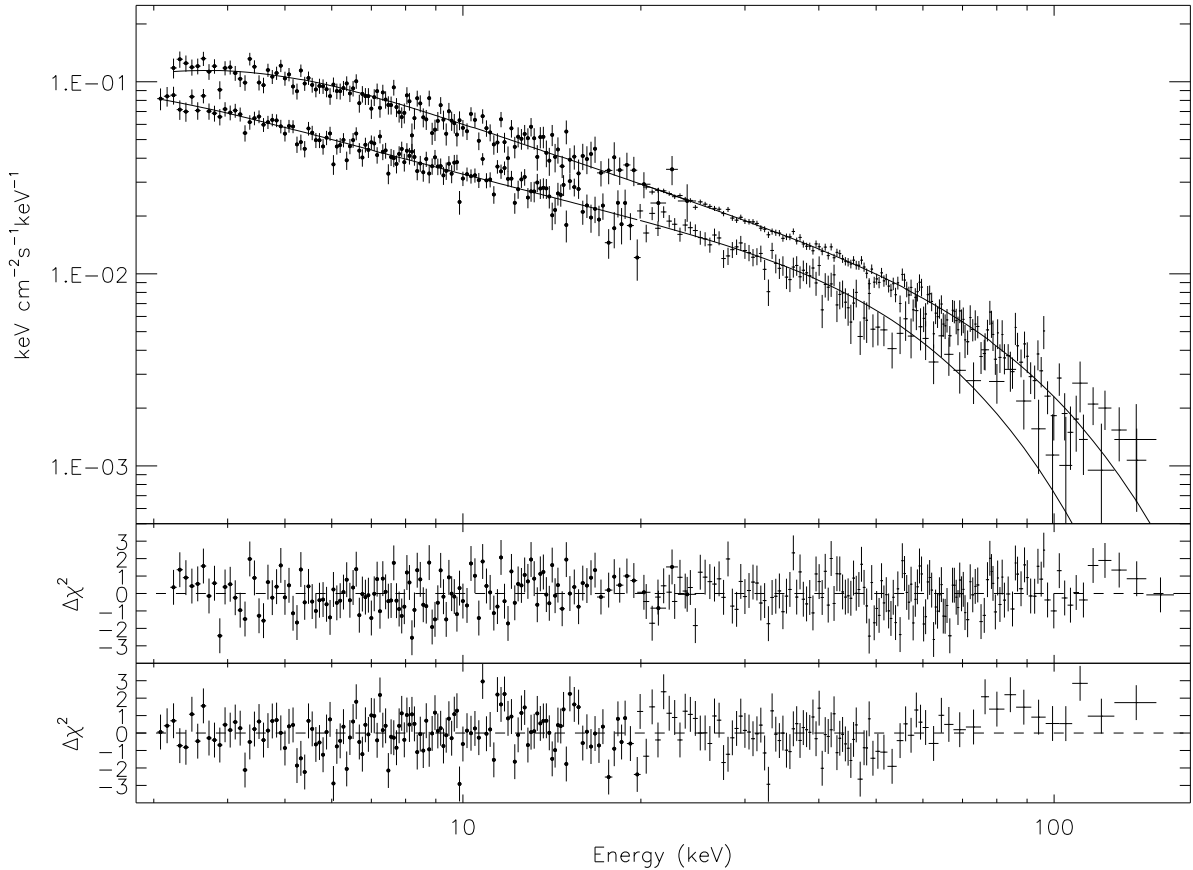}
      \caption{The \INTEGRAL spectrum of \GS as observed in Autumn 2003 (brighter spectrum) and Autumn 2006.
      Simple crosses and dotted crosses are used for ISGRI and JEM-X, respectively.
      Residuals to the best \CompTB fits are shown in the lower panels for the
      2003 (above) and 2006 data (below).
              }
         \label{Fig1}
   \end{figure*}

\section{Spectroscopic results}
Our wide band \INTEGRAL spectra of \GS can be satisfactorily fit in a first approximation by a simple, empirical, 
exponentially cutoff power-law model, \XSPEC's {\sc cutoffpl} (see Table 2). The power-law photon index is $\sim 1.6$ and the cutoff energy is $\sim 50$ keV in all the observations. 
Such values are in good agreement with all the previous results based on this
simple model, thus suggesting that the overall spectral shape (and so the spectral state) of the source is almost constant with time on a time scale of several years.

From a less empirical, more physical point of view, a single high
energy cutoff power law is a clear indication of a single Compton region at
work. This is (at least partially) in agreement with most of the previous
  results, but not with those obtained by Thompson et al. (2005) 
(see also \cite{Thompson2}), who included a second, distinct Compton component
with a different geometry and electron temperature to fit their joint
\Chandra/\XTE data:~in our case, as {\sc cutoffpl} is an acceptable model,
there is a clear indication that the data themselves do not require the use of too many free parameters to explain the observed emission from \GS.

Taking all this into account, we tried to fit the data with a single thermal Compton component. The recently 
developed thermal plus bulk Comptonisation model (\cite{Farinelli}), \CompTB, included in the standard \XSPEC distribution\footnote{http://heasarc.nasa.gov/docs/xanadu/xspec/newmodels.html}, is presently 
the most up-to-date Comptonisation model in a diffusion regime (such approximation holds well for NS-LMXBs). 
\CompTB is a general model including, in a self-consistent way, the seed
photons emission together with the Comptonised emission, whether the latter is
either purely thermal (as in the well-known model \CompTT), or mixed dynamical (bulk Comptonisation) plus thermal.
As a detailed description of the model is given in Farinelli et al. (2008), 
we here only point out that the free parameters of \CompTB are the blackbody seed photon 
and the electron temperatures (kTs and kTe, respectively), the Comptonisation 
index $\alpha$, the bulk parameter $\delta$, the covering factor log(A) and the 
normalisation.

As already mentioned above, the spectrum of \GS is well described by a  
simple function ({\sc cutoffpl}), which mimics unsaturated Comptonisation for static 
plasma and no blackbody-like peak is observed at low energies: 
the case of the seed blackbody radiation dominated by the Compton emission is modeled by \CompTB with a high value of the covering factor, i.e. $A>>1$.
Moreover, \GS being a low luminosity (a few \% Eddington) NS-LMXB in a
  low (hard) state, the detection of a bulk emission is unlikely (\cite{Paizis}, see also Sect. 4.1). 
These two particular cases are well reproduced in our fits: when allowed to
vary, both the log(A) and the $\delta$ parameters are eventually pegged by
the fitting procedure to their hard limits corresponding to total
  coverage (log(A)$=8$) and no bulk emission ($\delta=0$), thus reproducing a simple thermal Comptonisation spectrum without direct emergence of the underlying seed photons.
We then fixed $\delta$ and log(A) to the above limit values in all the fits, leaving only kTs, kTe and $\alpha$ as variable  parameters.
In this particular case, the dimension of the free-parameter space is identical to that of the well consolidated and widely adopted \CompTT model (\cite{Titarchuk}), and the two models are substantially equivalent.

The results of our \CompTB fits are reported in Table 2 for each of the three different observation periods.
Spectra and residuals to the best fits are also shown in Fig. 1 for two selected observations.
A single \CompTB component with $kT_{e}$ in the range 15--20 keV effectively describes the shape of the spectra. 
Though the source persisted in its low-hard state, significantly different best fit values were obtained for both electron plasma and soft seed photon temperatures in the 2003 and 2006 observations. 
The 2006 data set required lower temperatures and a lower value of the index
$\alpha$, indicating an increased efficiency of the Compton upscattering. The
2--200 keV intensity also dropped by $\sim 30\%$ with respect to 2003.
We carefully verified that the different values of the main spectral
  parameters in 2006 are not biased by a probable long term variation of the
  JEM-X/ISGRI normalisation constant: in fact consistent $kT_{e}$ values are
  obtained even when forcing the 2003 cross-calibration in the 2006 data set,
  and vice versa. Moreover, considering only ISGRI data for the fits does
  not alter the main parameter values. Finally, the \XTE/ASM data
independently confirm in a narrower passband the drop in the flux (see
  Table 1).

Best-fit excesses above $\sim 75$ keV are observed in the 2006 data set (see
  lower panel of Fig. 1) and partly in the Autumn 2003 one. But they are not detected
  in Spring 2003 and moreover their significance is poor. So we regard such
  residuals as random fluctuations. 

Interesting additional physical information can be extracted from the best-fit parameters of \CompTB; for example, the cloud optical depth $\tau$ can be derived by the eqs. [17]-[24] in \cite{Titarchuk2}. The $\tau$ values, assuming a spherical geometry, are included in Table 2.  
For a direct comparison, the results obtained replacing \CompTB with the more popular model \CompTT (spherical geometry assumed) are also reported in Table 2. As expected, the main \CompTT parameter values are in good agreement with the \CompTB ones.

Moreover, the order of magnitude of the radius of the (assumed) spherical region emitting the blackbody seed photons can be estimated: as the normalisation in \CompTB is defined as $C_{\rm N}=L_{\rm 39}/D^2_{10}$, 
where $L_{39}$ is the seed photon spectrum luminosity (in units of $10^{39}$
erg ${\rm s}^{-1}$) and $D_{10}$ is the source distance (in units of 10 kpc),
one can derive applying the Stefan-Boltzmann law the emitting region radius as
\begin{equation}
R_{\rm km}= 0.88\times 10^{2} \frac{C_{\rm N}^{1/2}~D_{10}}{kT_{\rm s}^2}
\label{radius}
\end{equation}
We obtain for the three observational data sets the values $R_{\rm km}=$ 6,~8
and 24 km, respectively (a source distance of 7 kpc is assumed).

The X-ray energy amplification caused by the Compton scattering can also be determined by computing the Compton Amplification Factor ($CAF$), defined as the ratio of the Comptonised bolometric energy flux to that of the blackbody seed radiation.
The obtained values of the parameter for each observation of \GS are also listed in Table 2. Increased Compton efficiency, as already testified by the correspondingly decreasing value of the $\alpha$ index, is observed in 2006.

   \begin{figure}
   \includegraphics[width=8.8cm]{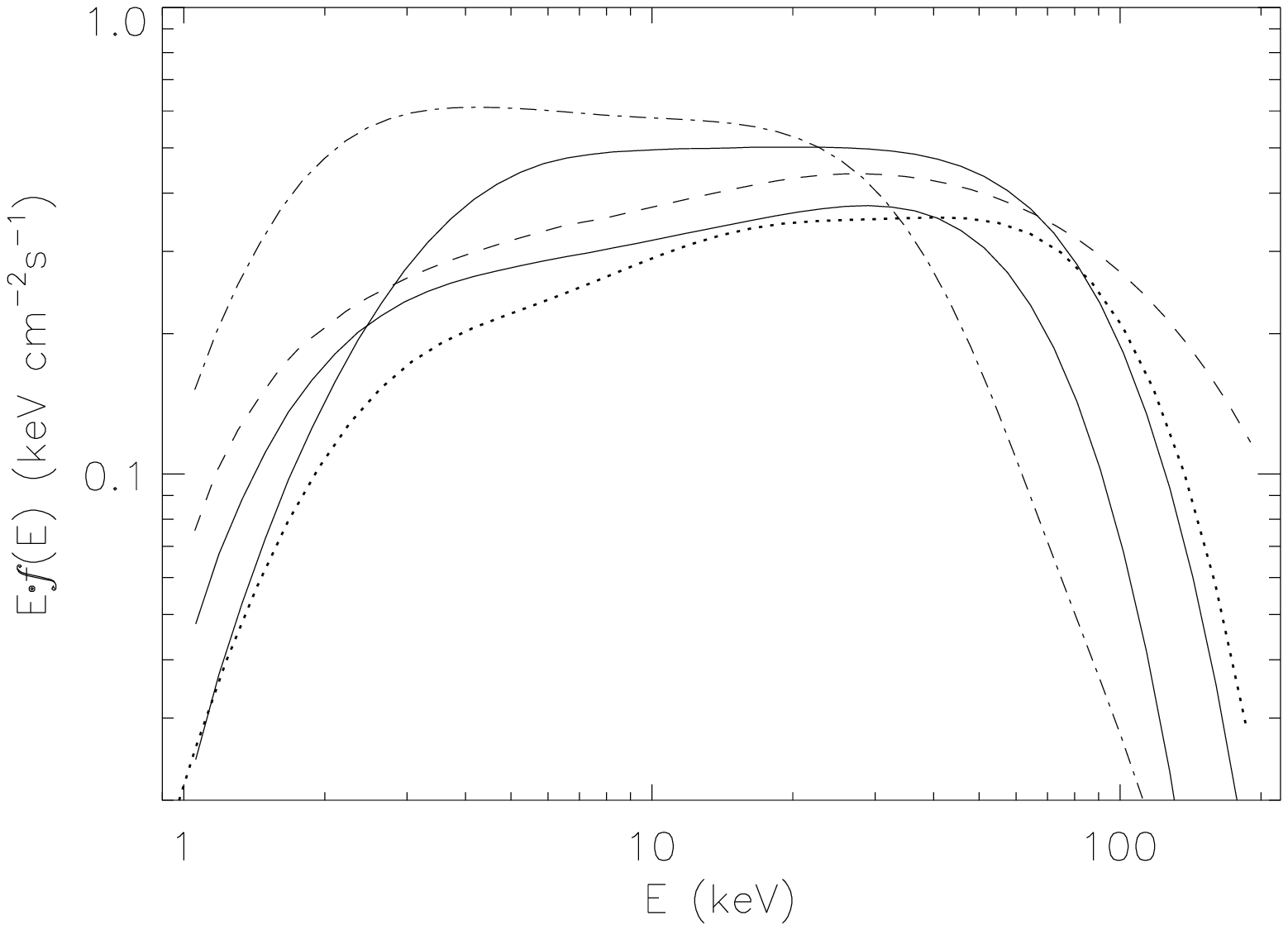}
      \caption{Comparison of wide band spectra of \GS obtained by different
        observations: 
      our \INTEGRAL spectra, as in Figure 1 (simple lines);
      \XTE (\cite{Barret}, dashed line);
      \Chandra$+$\XTE (\cite{Thompson}, dash-dotted);
      \SAX (\cite{Zand}, dots).
      The Fe lines, even when detected, are not included.
                    }
         \label{Fig2}
   \end{figure}

\section{Discussion}
\subsection{\GS in a unifying framework}
We applied a recently developed model for Compton upscattering, \CompTB,
  to study the low-hard state variability of \GS. 
As Paizis et al. (2006) demonstrated, 
general Comptonisation models including bulk emission are very effective in
modeling the variety of NS-LMXBs spectral states and the state transitions in
individual sources. This is obtained, in a natural way, by smooth changes in the
relative importance of bulk over thermal Comptonisation.
In this scenario having the interplay of bulk and thermal Compton emission as
unifying parameter, a low-hard state is expected to originate at low accretion rates, where the coronal Compton cooling is far from being efficient, thus leaving out a {\em hot screen} which extends radially and spherically, thus shielding the NS surface and allowing the direct accretion disk emission to be cooler.  
Bulk motion Comptonisation can still occur in this case, as the NS gravity is
not balanced by the radiation pressure, but the effect is expected to be weak
and probably washed in the hot screen with negligible effects for a distant 
observer. 

The present results confirm the above picture for the case of \GS, as pure thermal upscattering  
completely accounts for the source emission.  
The relative extension of the Compton cloud determines whether it is possible to observe a fraction of the seed photons' population originating from the NS surface and/or from the inner accretion disk. 
In our spectra the soft blackbody seed photons are not directly detected, given the inferred $\log~A\gg 1$, and no further soft components are needed to fit the data.

\subsection{Wide band spectroscopy: extra spectral components and variability}
Undetection of additional soft components in the \INTEGRAL data is in
agreement with the results obtained by Strickman
  et al. (1996) and Barret et al. (2000)  
for the same source, but not with those of in 't Zand et al. (1999) and Del Sordo et al. (1999),
who proposed an extra blackbody ($k{\rm T}_{bb}$ of $\sim 3$ keV and $\sim 1$ keV, respectively), in addition to a thermally Comptonised component (or a Compton-like, in the case of \cite{DelSordo}), to fit the \SAX data. 
Our upper limits for a further blackbody-like component were calculated,
  assuming a typical colour temperature of 1 keV and a distance of 7 kpc. We
  obtained blackbody emitting regions with radii R$<1.0$ km and R $<0.7$ km for
  the 2003 and 2006 data sets, respectively (90\% confidence).  

Applying a {\sc pexrav} model to our data as in Barret et al. (2000)
  does not improve the fits. The simple cutoff power law is reproduced, but
  the reflection parameter is pegged to zero. This is expected, as {\sc
  cutoffpl} always yields acceptable fits and there is no need for further
hypotheses.

For \GS, Thompson et al. (2005) 
and (2008) proposed a model with two distinct Comptonising regions, one of spherical 
shape and a high ($\sim 20$ keV) temperature, the other with disk geometry and
a lower ($\sim 5$ keV) temperature. The two separate regions had different opacities and seed photons populations.
This two-cloud model actually fits our data, with values very close to those obtained by the above authors, but this is not very surprising because it has many free parameters.
As a matter of fact, our data do not need too many free parameters because a
simple cutoff powerlaw provides a satisfactory fit. Moreover, a single powerlaw
is an indication of a single Comptonising region at work. 
In \cite{Thompson}, much simpler models (e.g. {\sc cutoffpl}) yielded acceptable fits to
their data as well; interestingly, all the fits had very low ($\sim 0.7$--$0.8$)
reduced chisquared, probably indicating systematic effects at work, such as error overestimation or missed rebinning.

It must be stressed that the presence of a second, softer, spectral component cannot be completely ruled out by the present \INTEGRAL data. But its actual need should be justified by the overall X-ray spectral shape obtained including the lowest energy range (E$<$ 2--3 keV), which is not covered by JEM-X. 
This is likely, as both the \SAX and \Chandra already published spectra start well below 1 keV.
Nevertheless, both the previously reported $\sim 3$ keV blackbody-like
component (\cite{Zand}) and the $\sim 5$ keV Comptonised emission (\cite{Thompson}) are well inside the \INTEGRAL passband and should be unambiguously constrained. This is not the case.

The present analysis shows clear evidence for intensity and spectral
variability of \GS (even though almost unexpected, given the peculiar long
time-scale regularity of its bursting behaviour, which is suggestive of a stable, smooth accretion).
In particular, significantly lower plasma temperatures and 2--200 keV
luminosity are observed in 2006 compared with 2003.

A comparison of our results with previously reported wide band
  observations of \GS is shown in Fig. 2. While at least approximately the canonical low-state spectral
  hardening with source dropping luminosity is well reproduced by our
  2003 data, the unusually low electron plasma temperature of 2006 coupled
  with relatively low luminosity seems to contradict this consolidated scenario.
Together with the probably erratic presence of blackbody-like emission at
lower energies (see also below), the overall picture of \GS is perhaps
suggestive of the emergence of possible sub-states within its low-hard
state. In such a spectral state, different observational characteristics could be attributed to differences in the configuration of the accreting system, with the size and geometry of the Compton cloud likely playing a pivotal role.

\subsection{X-ray bursts timing irregularities}
Taking into account the observed decrease of the X--ray bursts' wait times
together with lower measured intensity (see Table 1), our November 2006 data
show a clear deviation from the typical limit-cycle X-ray bursting behaviour
of \GS. Burst wait times $\Delta$T are generally anticorrelated with very good approximation to the
source intensity $F_{X}$ by the simple formula 
$\Delta$T $\propto F_{X}^{-1.05}$ 
(see \cite{Galloway}).
If one were to consider the measured 2--200 keV flux as an indication of the
actual accretion rate as well as the burst fuel accumulation rate, a deviation
of $\sim 35\%$ from the limit-cycle rule would be obtained with respect to e.g. the 2003 \INTEGRAL data.
Similar anomalies were already observed in the April 2003 \XTE data
(\cite{Thompson2}). These results indicate that spectral measurements which do
not include the very soft X-rays are unreliable tools to extrapolate the bolometric luminosity, which can be widely underestimated.  
In fact Thompson et al. (2008) 
reported of a simultaneous XMM-observed transient blackbody-like spectral component. This very soft (a few tenths of keV) component accounted for the missing flux, allowing for a better estimate of the bolometric luminosity and so of the true accretion rate.
Our \INTEGRAL results can also be explained in the same natural way, leaving out early burst ignition and/or anisotropy considerations. But a further very soft component similar to the one reported by Thompson et al. (2008) 
would then remain undetected, being outside of the JEM-X passband.
Nevertheless, our 2006 data show a significant decrease of both the Compton cloud electron and the seed photon temperatures, which is possibly related to an overall bolometric spectral variation of \GS. 

The above reported erratic variability of the source, in terms of both transient soft components and/or Comptonisation parameters, deserves to be studied in detail to be better understood and implies the need for further accurate wide band (from very soft X-rays up to soft Gamma-rays) monitoring programs of \GS and other similar X-ray bursters.

\begin{acknowledgements}
This work has been supported by the grant from the INAF PRIN 2007 {\em Bulk motion Comptonization models in X-ray Binaries: from phenomenology to physics}, PI M. Cocchi.

A.P. acknowledges the Italian Space Agency financial and programmatic support via contract I/008/07/0.

The authors thank C. Ferrigno and A. Segreto for useful suggestions and tips.
\end{acknowledgements}


\begin{thebibliography}{}

   \bibitem[Barret, Motch \& Pietsch, 1995]{Barret95} Barret, D., Motch, C., and Pietsch, W., 1995, \aap, 305, 526

   \bibitem[Barret et al., 2000]{Barret} Barret, D., Olive, J.F., Boirin, L., Done, C., Skinner, G.K., \& Grindlay, J.E., 2000, \apj, 533, 329
   
   \bibitem[Cocchi et al., 2001]{Cocchi} Cocchi, M., Bazzano, A., Natalucci, L., et al., 2001, Advances in Space Research, 28, 375

   \bibitem[Cornelisse et al., 2003]{Cornelisse} Cornelisse, R., in~'t~Zand, J.J.M., Verbunt, F., et al., 2003, \aap, 405, 1033
   
   \bibitem[Del Sordo et al., 1999]{DelSordo} Del Sordo, S., Frontera, F., Pian, E., et al., 1999, Astrophys. Lett. and Comm., 38, 125

   \bibitem[Dickey \& Lockman, 1990]{Dickey} Dickey, J., \& Lockman, F. 1990, ARA\&A, 28, 215

   \bibitem[Farinelli et al., 2008]{Farinelli} Farinelli, R., Titarchuk, L., Paizis, A., \& Frontera, F., 2008, \apj, 680, 602

   \bibitem[Galloway et al., 2004]{Galloway} Galloway, D.K., Cummings, A., Kuulkers, E., Bildsten, L., Chakrabarty, D., \& Rothschild, R.E. 2004, \apj, 601, 466

   \bibitem[Homer, Charles \& O'Donoughue, 1998]{Homer} Homer, L., Charles, P.A., and O'Donoghue, D., 1998, MNRAS, 298, 497

   \bibitem[Lebrun et al., 2003]{Isgri} Lebrun, F., Leray, J.P., Lavocat, P., et al., 2003, \aap, 411, L141

   \bibitem[Lund et al., 2003]{JemX} Lund, N., Budtz-Jørgensen, C., Westergaard, N.J., et al., 2003, \aap, 411, L231

   \bibitem[Makino et al., 1988]{Makino} Makino, R., et al., 1988, IAU Circ., 4653

	 \bibitem[Mineo et al., 2006]{Mineo} Mineo, T., Ferrigno, C., Foschini, L., et al., 2006, \aap, in press (arXiv:astro-ph/0601641v1)

   \bibitem[Paizis et al., 2006]{Paizis} Paizis, A., Farinelli, R., Titarchuk, L., et al., 2006, \aap, 459, 187

   \bibitem[Segreto \& Ferrigno, 2006]{Segreto} Segreto, A., \& Ferrigno, C., 2006, in Proc. of the 6th \INTEGRAL Workshop {\em The Obscured Universe}, Moscow on 2--8 July 2006, ed. S. Grebenev, R. Sunyaev, \& C. Winkler, p.633

   \bibitem[Strickman et al., 1996]{Strickman} Strickman, M., Skibo, J.,
    Purcell, W., Barret, D., and Motch, C., 1996, A\&AS, 120, 217

   \bibitem[Thompson et al., 2005]{Thompson} Thompson, T.W.J., Rothschild, R.E., Tomsick, J.A., \& Marshall, H.L., 2005, \apj, 634, 1261

   \bibitem[Thompson et al., 2008]{Thompson2} Thompson, T.W.J., Galloway, D.K., Rothschild, R.E., \& Homer, L., 2008, \apj, 681, 506
   
   \bibitem[Titarchuk, 1994]{Titarchuk} Titarchuk, L., 1994, \apj, 434, 570

   \bibitem[Titarchuk \& Lyubarskij (1995)]{Titarchuk2} Titarchuk, L., \& Lyubarskij, Y., 1995, \apj, 450, 876

   \bibitem[Ubertini et al., 1997]{Ubertini97} Ubertini, P., Bazzano, A., Cocchi, M., et al., IAU circ. 6611, 1997.

   \bibitem[Ubertini et al., 1999]{Ubertini99} Ubertini, P., Bazzano, A., Cocchi, M., et al., {\em ApJ}, 514, L27-L30, 1999.

   \bibitem[Winkler et al., 2003]{Integral} Winkler, C., Courvoisier, T. J.-L., Di Cocco G., et al. 2003, \aap, 411, L1

   \bibitem[in~'t~Zand, 1992]{Zand92} in~'t~Zand, J.J.M., 1992, Ph.D. thesis, University of Utrecht

   \bibitem[in~'t~Zand et al., 1999]{Zand} in~'t~Zand, J.J.M., Heise, J., Kuulkers, E., Bazzano, A., Cocchi, M., \& Ubertini, P., 1999, \aap, 347, 891

\end{thebibliography}
\end{document}